\documentclass[fleqn,twoside,twocolumn,nofootinbib,showkeys]{revtex4} 
\usepackage[nocpr,nopacs]{ujp} 

\usepackage{graphicx,epic}
\usepackage{epstopdf}
\usepackage{float}
\usepackage[percent]{overpic}

\newcommand{\Tr}{\mathop{\mathrm{Tr}}\nolimits}
\newcommand{\diag}{\mathop{\mathrm{diag}}\nolimits}

\begin{document}
\title[Composite Fermions  
        as Deformed Oscillators: Wavefunctions and Entanglement]
      {COMPOSITE FERMIONS\\ 
        AS DEFORMED OSCILLATORS:\\ WAVEFUNCTIONS AND ENTANGLEMENT\hspace*{0.4mm}$^1$}%
\author{A.M.~Gavrilik}
\affiliation{Bogolyubov Institute for Theoretical Physics, Nat. Acad. of Sci. of Ukraine}
\address{14b, Metrolohichna Str., Kyiv 03143, Ukraine}
\email{omgavr@bitp.kiev.ua}
\author{Yu.A.~Mishchenko}
\affiliation{Bogolyubov Institute for Theoretical Physics, Nat. Acad. of Sci. of Ukraine}
\address{14b, Metrolohichna Str., Kyiv 03143, Ukraine}

\udk{530.145, 539.12.01} \razd{\seci}

\autorcol{A.M.\hspace*{0.7mm}Gavrilik, Yu.A.\hspace*{0.7mm}Mishchenko}%

\setcounter{page}{1134}%

\begin{abstract}
Composite structure of particles somewhat modifies their statistics,
compared to the pure Bose- or Fermi-ones.\,\,The spin-statistics
theorem, so, is not valid anymore.\,\,Say, $\pi$-mesons, excitons,
Cooper pairs are not ideal bosons, and, likewise, baryons are not
pure fermions.\,\,In our preceding papers,
we studied bipartite \textit{composite boson} (i.e.\,\,quasiboson)
systems via a realization  by deformed oscillators.\,\,The\-rein,
the interconstituent entanglement characteristics such as
entanglement entropy and purity were  found in terms of the
parameter of deformation.\,\,Herein, we perform an analogous study
of composite \textit{Fermi-type} particles, and explore them in two
major cases: (i) ``boson + fermion'' composite fermions (or
cofermions, or CFs); (ii) ``deformed boson~+ fermion'' CFs.\,\,As we
 show, cofermions in both cases admit only the realization by
ordinary fermions.\,\,Case~(i) is solved explicitly, and admissible
wavefunctions are found along with entanglement
measures.\,\,Case~(ii) is treated within few modes both for CFs and
constituents.\,\,The entanglement entropy and purity of CFs are
obtained via the relevant parameters and illustrated graphically.
\end{abstract}


\maketitle

\section{Introduction}

Composite fermions play a significant role in modern quantum
physics.\,\,Suf\-fice it to mention a few instances of CFs: these
include quasiparticles involved in the theory of fractional quantum
Hall effect~\cite{Jain2007Composite}; also, let us mention baryons
and pentaquarks as known instances in hadron
physics~\cite{Hadjimichef1998,Oh2004Penta,Browder2004Penta}.\,\,In
this paper, we will focus on the entanglement properties of
composite
  fermions basing on their fermionic oscillator realization,
  in two relatively simple cases: cofermion
  {\it built from} a pure fermion and a pure boson, or the cofermion as
  a composite {\it made of} a pure fermion and
  a deformed boson (taken in general form).


    In our preceding works~\cite{GKM2,GKM,GM_Entang,GM_Ent(En)}, we
studied bipartite (two-component) composite ``bosons'' of two types:
``fermion~+ fermion'' and ``boson~+ boson'' with creation and
annihilation operators within the typical ansatz
\begin{equation}\label{ansatz}
A^\dag_\alpha= \sum\limits_{\mu\nu} \Phi_\alpha^{\mu\nu} a^\dag_\mu
b^\dag_\nu,\ \quad A_\alpha = \sum\limits_{\mu\nu}
\overline{\Phi_\alpha^{\mu\nu}} b_\nu a_\mu,
\end{equation}
where the creation operators~$a^\dag_\mu$, $b^\dag_\nu$ for the
(distinguishable) constituents taken either both fermionic or both
bosonic.\,\,In~\cite{GKM2,GKM}, it was shown that composite bosons
 of a particular form (with appropriate wavefunctions~$\Phi_\alpha^{\mu\nu}$)
 can be realized, in the operator sense, by suitable deformed bosons
 (deformed \mbox{oscillators}).\looseness=1


\footnotetext[1]
           {This work is based on the results presented at the
           XI Bolyai--Gauss--Lobachevskii (BGL-2019) Conference: Non--Eucli\-de\-an,
           Noncommutative Geometry and Quantum Physics.}An important concept in quantum information theory, quantum communication,
and teleportation~\cite{Horodecki,Tichy_Rev} is the notion of
entanglement or quantum correlation between the constituents of a
composite particle or composite system.\,\,Re\-cent\-ly, this
concept was actively studied just in the context of composite
bosons~\cite{Law,Chudzicki,GM_Entang, Lasmar2019Asm2Nferm}.\,\,Among
the measures characterizing the degree of entanglement, best known
are the entanglement entropy and purity (= inverse of the Schmidt
number)~\cite{Horodecki,Tichy_Rev}.\,\,The measures of
intercomponent entanglement in a quasiboson quantify to what extent
the properties of a quasiboson approach those of a true
boson~\cite{Law,Chudzicki,Ramanathan,Morimae}.

 For composite bosons realizable by {\it deformed} quantum oscillators, it is
possible  to directly link~\cite{GM_Entang} the relevant {\it
parameter of deformation} with the entanglement characteristics of a
composite boson.\,\,Then the characteristics (or measures) of
bipartite entanglement with respect to $a$- and $b$-subsystems,
see~(\ref{ansatz}), can be found explicitly~\cite{GM_Entang} in
terms of the deformation parameter: for a single composite boson,
for multi-quasiboson states, and for coherent states constructed for
such quasibosons.

As a very important issue, the influence of system's energy on
quantum correlation and/or quantum statistics properties of the
system was studied, for the quasibosons,
in~\cite{GM_Ent(En)}.\,\,The energy of the quasiboson differs from
the energy of a respective ideal boson, and the difference
(including the energy of bound states) essentially
depends~\cite{GM_Ent(En)} on the quasiboson's entanglement, clearly
showing a deviation from the bosonic behavior.\,\,Such
entanglement-energy relation is relevant to quantum information
research, quantum communication, entanglement production
\cite{Weder}, quantum dissociation processes \cite{Esquivel},
particle addition or subtraction~\cite{Kurzynski,Bartley},
\textit{etc}.

Below, we explore the cofermions.\,\,Since the  entanglement entropy
is of primary interest, after the appropriate analysis of the
realization issue, we find the entanglement
entropy~$S_{\mathrm{ent}}$ characterizing the composite
fermion.\,\,Our treatment is performed for one-CF states (to
compare, the respective results for one  quasiboson states are also
briefly sketched).\,\,In some analogy with the case of composite
bosons, we take the cofermions as bipartite systems realized by
mode-independent\,\footnote[2]{This is understood in the fermionic,
i.e.\,\,anticommuting, sense.} fermionic oscillators.\,\,Ano\-ther
entanglement measure, purity, is considered in a special~case.

We have to emphasize that our investigation of entanglement concerns
not a many-cofermion system, but the states of a single (or
isolated) cofermion.\,\,Ac\-cor\-ding\-ly, the considered
entanglement and its entropy concern two constituents of the
bipartite CF. These features make our approach and analysis
essentially different from some recent works on the entanglement
entropy of a system of free or composite fermions, see,
e.g.,~\cite{Gioev2006Entanglement,Shao2014Entanglement}, where the
spatial size of a subsystem plays the basic role, and entanglement
entropy \textit{was viewed in a way different from our~one}.

In Sec.~\ref{sec:setup}, a sketch of the realized composite bosons
is given.\,\,Sections~\ref{sec:QFs}--\ref{sec:2mod} deal with
cofermions: we perform the analysis of operator-level realization of
cofermions by (deformed) fermionic oscillators.\,\,Then the
entanglement entropy of the CF one-particle states is
explored.\,\,Mo\-di\-fied CFs (composed of fermion and deformed
boson) are analyzed in Sec.~\ref{sec:2mod}(b).\,\,The purity for a
CF state is considered in Sec.~\ref{sec:2mod}(a).\,\,The paper ends
with conclusions.\vspace*{-1mm}


\section{Quasibosons Formed\\ as Two-Fermion (Two-Boson) Composites~\cite{GKM2}}\label{sec:setup}

 Recall the main facts about composite bosons realized~\cite{GKM,GKM2}
by a set of independent modes of deformed bosons (deformed
oscillators), given by the defining {\it deformation structure
function}~$\varphi(n)$~\cite{Meljanac}.\,\,At the algebraic level,
the quasiboson operators~$A_\alpha$, $A^\dag_\alpha$ and the number
operator~$N_\alpha$ satisfy the same relations {\it on the states}
as the corresponding deformed oscillator creation/annihilation and
number operators:\vspace*{-1mm}
\[
A^\dag_\alpha A_{\alpha} \!\simeq\! \varphi(N_\alpha),~~~
[A_\alpha,A^\dag_\beta] \!\simeq\! \delta_{\alpha\beta}
\bigl(\varphi(N_\alpha \!+\!1) \!-\! \varphi(N_\alpha)\bigr),
\]\vspace*{-9mm}
\[
[N_\alpha,A^\dag_\beta] \simeq \delta_{\alpha\beta}
A^\dag_\beta,\quad [N_\alpha,A_\beta] \simeq - \delta_{\alpha\beta}
A_\beta .
\]
Here, $\simeq$ denotes weak equality (i.e. on the states), symbols
$\delta_{\alpha\beta}$ reflect mode independence.\,\,In such
realization, the structure function~$\varphi(n)$
involves~\cite{GKM,GKM2} the discrete deformation parameter~$f$ and
is quadratic in the quasiparticles number~$n$ \ ($\kappa= +1$ or
$-1$ for two bosonic or two fermionic constituents):\vspace*{-1mm}
\begin{equation}\label{phi(n)}
\varphi(n)=\Biggl(\!1+\kappa \frac{f}{2}\!\Biggr)n - \kappa
\frac{f}{2}n^2,\quad f=\frac{2}{m},
\end{equation}
$m=1,2,...$\,.
 The matrices $\Phi_\alpha$ in~(\ref{ansatz}) are~\cite{GKM,GKM2}\vspace*{-1mm}
\begin{equation}
\Phi_\alpha = U_1(d_a) \diag\Bigl\{0..0,\sqrt{f\!/2}\,
U_\alpha(m),0..0\Bigr\} U^\dag_2(d_b),\!\!\! \label{gen_solution}
\end{equation}
where $U_j(r)$ stands for an arbitrary unitary $r\times r$ matrices,
the dimension $d_a$ or $d_b$ is the total number of modes
for constituents with operators $a_\mu$ or $b_\nu$.

Note that the state of one composite boson,\vspace*{-1mm}
\[
|\Psi_\alpha\rangle \!=\! \sum\limits_{\mu\nu} \Phi_\alpha^{\mu\nu}
|\mu\rangle \!\otimes |\nu\rangle,\quad |\mu\rangle \equiv
a^\dag_\mu|0\rangle,\ \ \ |\nu\rangle \equiv b^\dag_\nu|0\rangle,
\]
is, in general, bipartite-entangled relative to the states of two
constituent fermions (or two bosons). There are well-known measures
of entanglement~\cite{Tichy_Rev,Horodecki}: Schmidt rank~$s$,
Schmidt number~$K$ or its inverse~-- purity, entanglement
entropy~$S$, and concurrence~$C$.\,\,As proven in~\cite{GM_Entang},
the (internal) entanglement entropy of one composite
boson\vspace*{-1mm}
\begin{equation}
S_{\rm ent} = \ln(m) = \ln(2/f) .
\end{equation}\vspace*{-5mm}

\noindent For the multi-quasibosonic states, the respective extended
results were also derived, see~\cite{GM_Entang,GM_Ent(En)}.

The other known measure of entanglement~\cite{Tichy_Rev,Horodecki},
purity is inverse to the Schmidt number: $P=1/K$. Note that {\it
purity} was exploited in connection with the issue of entanglement
creation using scattering processes~\cite{Weder} (for other
contexts, see~\cite{Kurzynski,McHugh}).\,\,For {\it one-quasiboson
entangled system}, purity is as well linked~\cite{GM_Entang} with
the deformation parameter $m=\frac2f$:\vspace*{-1mm}
\begin{equation}\label{pur_def}
P\!=\! \sum\limits_k \lambda_k^4 = \frac1m
\end{equation}\vspace*{-3mm}

\noindent and thus takes discrete values.\vspace*{-1mm}

\section{Cofermion -- as Fermion\\ \textit{plus} Deformed Boson}\label{sec:QFs}

From now on, we consider composite fermions which are composed of a
pure (or deformed) boson and a pure fermion, from the viewpoint of
their realization by deformed fermions.\,\,Like for composite
bosons, the realization of CFs is required to be constructed in the
way that enables to treat CFs' creation, annihilation, and number
operators as the respective operators of deformed fermions.\,\,In
turn, this gives the advantages compared to the standard quantum
mechanical or macroscopic approach with explicit direct
consideration of a composite structure,~-- to perform calculations,
in the effective or approximate description.\,\,CFs' creation and
annihilation operators are given by the same ``ansatz'' as
in~(\ref{ansatz}), but now $a^{\dag}_{\mu}, a_{\mu}$ are,
respectively, the creation and annihilation operators for the
constituent bosons (deformed or not) and $b^{\dag}_{\nu},
b_{\nu}$~-- those for the constituent fermions, with standard
anticommutation relations for the latter.\,\,For simplicity, we
suppose that different modes of deformed bosons are
independent.\,\,Then we have the following defining commutation
relations for the operators of constituent bosons (deformed or not;
$n^a_\mu$ is the particle number operator for deformed bosons in the
$\mu$-mode):\vspace*{-1mm}
\[
a^\dag_\mu a_\mu=\chi(n^a_\mu),~~~~[a_\mu,a^\dag_{\mu'}] =
\delta_{\mu\mu'} \bigl(\chi(n^a_\mu\!+\!1) \!-\!
        \chi(n^a_\mu)\bigr);
        \]\vspace*{-9mm}
        \[
[a^\dag_\mu,a^\dag_{\mu'}]=0;~~~ [n^a_\mu,a^\dag_\mu]=a^\dag_\mu.
\]
Here, the deformation structure function $\chi(n)$ means the general
case of deformed constituent boson.\,\,For a non-deformed,
i.e.~usual, boson, $\chi(N)\equiv N$.\,\,In the case of deformation,
$\chi(N)$ depends on one or more deformation parameters which admit
a ``no-deformation'' limit.\,\,It  is convenient to work with
deformed boson states normalized as $|\mu\rangle=a^{\dag}_{\mu}
|0\rangle$.\,\,The same concerns CF states, obeying the
normalization condition for structural matrices
(wavefunctions):\vspace*{-1mm}
\begin{equation}  \label{orthonorm}
\Tr(\Phi_\beta \Phi^\dag_\alpha) = \delta_{\alpha\beta}.
\end{equation}

 We suppose CFs to behave themselves on the states as deformed particles with
a structure function $\varphi(N)$.\,\,Res\-pective deformed fermions
providing the realization are supposed to be independent (in the
fermionic sense).\,\,De\-fining the number operator $N_\alpha$ for
CFs as $\varphi(N_\alpha) \simeq A^\dag_\alpha A_\alpha$, we write
the realization conditions:\vspace*{-1mm}
\begin{equation}
\{A_\alpha,A^\dag_\beta\} \simeq \delta_{\alpha\beta}
    [\varphi(N_\alpha\!+\!1) \!+\! \varphi(N_\alpha)], \label{req2'}
\end{equation}\vspace*{-9mm}
\begin{equation}
\{A^\dag_\alpha,A^\dag_\beta\}\simeq0,~~~\alpha\ne \beta; \quad
[N_\alpha,A^\dag_\beta]\simeq \delta_{\alpha\beta} A^\dag_\beta.
\label{req3}
\end{equation}
The first requirement in~(\ref{req3}) holds automatically, and
moreover, in the strict sense for all $\alpha$ and
$\beta$,\vspace*{-1mm}
\begin{equation}
\{A^{\dag}_{\alpha},A^{\dag}_{\beta}\} = 0,\ \text{particularly}\
(A^{\dag}_{\alpha})^2=0.        \label{nilpot}
\end{equation}
The latter, fermionic nilpotency, and~(\ref{req2'}) considered on
vacuum and one-CF states, along with the second equation
in~(\ref{req3}), yield the usual {\it fermionic structure function}
$\varphi(N)$:\vspace*{-1mm}
\begin{equation}
\varphi(0)=\varphi(2)=...=0,\quad \varphi(1)=\chi(1)=1.
\end{equation}
We analyze requirement~(\ref{req2'}), proceeding like
in~\cite{GKM2}, but alternate (interchange) the commutator and
anticommutator of the l.h.s.\,\,of (\ref{req2'})
with~$A_{\gamma_i}^\dag$.\,\,Introducing the notation\vspace*{-1mm}
\[
\Delta^{\!k}\chi(n^a_\mu) \!\equiv\! \sum\limits_{l=0}^k (-1)^{k-l}
{k\choose l} \chi(n^a_\mu\!+\!l), \ \ k\!=\!0,1,... ,
\]
with the first  terms\vspace*{-1mm}
\[
\Delta^0\chi(n^a_{\mu}) = \chi(n^a_{\mu}),\quad
 \Delta^1\chi(n^a_{\mu}) = \chi(n^a_{\mu}\!+\!1)-\chi(n^a_{\mu}),
 \]\vspace*{-9mm}
 \[
\Delta^2\chi(n^a_{\mu})=\chi(n^a_{\mu}\!+\!2)-2\chi(n^a_{\mu}\!+\!1)+\chi(n^a_{\mu}),
\]
for the anticommutator $\{A_\alpha,A^\dag_\beta\}$ in~(\ref{req2'}),
we have\vspace*{-1mm}
\[
\{A_\alpha, A_\beta^\dag\} = \sum\limits_\mu (\Phi_\beta
\Phi_\alpha^\dag)^{\mu\mu} \Delta^1\chi(n^a_\mu)\, +
\]
\[
+\sum_{\mu\mu'} (\Phi_\beta \Phi_\alpha^\dag)^{\mu'\mu}
a_{\mu'}^\dag a_\mu - \sum_{\mu\nu\nu'}
\overline{\Phi_\alpha^{\mu\nu}} \Phi_\beta^{\mu\nu'}\!
\Delta^1\chi(n^a_\mu)\, b_{\nu'}^\dag b_\nu.
\]
This will be exploited below.\,\,For a {\it nondeformed constituent
boson} ($\chi(n)\equiv n$), the latter with the use
of~(\ref{orthonorm}) reduces to\vspace*{-1mm}
\[
\bigl\{ \!A_\alpha, \!A_\beta^\dag \!\bigr\} \!=\! \delta_{\alpha\beta} + \!
\sum_{\mu\mu'}\! (\Phi_\beta\Phi^\dag_\alpha)^{\mu'\!\mu} a^\dag_{\mu'}a_\mu - \!
\sum_{\nu\nu'}\! (\Phi_\alpha^\dag\!\Phi_\beta)^{\nu\nu'}\! b^\dag_{\nu'}b_\nu.  
\]
Next, we calculate the commutator\vspace*{-1mm}
\[
\bigl[\{ A_\alpha, A^\dag_\beta\}, A^\dag_\gamma\bigr] \!=\!\!
\sum\limits_{\mu\mu_1\!\nu_1}\!\! \Bigl[(\Phi_\beta
\Phi^\dag_\alpha)^{\mu_1\mu}\Phi^{\mu\nu_1}_\gamma\, -
\]\vspace*{-5mm}
\[
-\, (\Phi_\gamma \Phi^\dag_\alpha)^{\mu_1\mu} \Phi^{\mu\nu_1}_\beta
\Bigr] a^\dag_{\mu_1} b^{\dag}_{\nu_1} \cdot
\bigl(\Delta^1\chi(n^a_\mu) +
\delta_{\mu\mu_1}\Delta^2\chi(n^a_\mu)\bigr)\, +
\]\vspace*{-6mm}
\begin{equation}
 +
\sum\limits_{\mu\nu\nu_1\!\nu_2} \overline{\Phi_\alpha^{\mu\nu}}
\Phi_\beta^{\mu\nu_1} \Phi_\gamma^{\mu\nu_2} a^\dag_\mu
b^\dag_{\nu_1} \Delta^2\chi(n^a_\mu)
b^\dag_{\nu_2}b_\nu.\label{commut1}
\end{equation}
A nondeformed analog (at $\chi(n)\equiv n$) of the latter
is\vspace*{-1mm}
\[
[\{A_\alpha,A^\dag_\beta\},A^\dag_\gamma] = \sum_{\mu\nu}
\bigl(\Phi_\beta \Phi^\dag_\alpha\Phi_\gamma - \Phi_\gamma
\Phi^\dag_\alpha \Phi_\beta\bigr)^{\mu\nu} a^\dag_\mu b^\dag_\nu.
\]
So, the validity of~(\ref{req2'}) on one-CF states $|\gamma\rangle$
yields the following relation of basic importance for the
wavefunctions:\vspace*{-1mm}
\[
(\Phi_\beta\Phi^\dag_\alpha\Phi_\gamma)^{\mu\nu} -
(\Phi_\gamma\Phi^\dag_\alpha\Phi_\beta)^{\mu\nu}\, +
\]\vspace*{-9mm}
\begin{equation} \label{req2_chi(2)}
+\, \bigl(\chi(2)\!-2\bigr)
\bigl[(\Phi_\beta\Phi^\dag_\alpha)^{\mu\mu} \Phi_\gamma^{\mu\nu}\!\!
-\! (\Phi_\gamma\Phi^\dag_\alpha)^{\mu\mu}\Phi^{\mu\nu}_\beta\bigr]
\!= 0. \,
\end{equation}
Note that, in the case of non-deformed constituent boson, this
relation yields $[\{A_\alpha,A^\dag_\beta\},A^\dag_\gamma] =
0$.\,\,That leads to a closed set of realization conditions on the
matrices $\Phi_\alpha$, see~(\ref{orthonorm}),(\ref{req2'}),
namely:\vspace*{-1mm}
\begin{equation}\label{nondef}
\Phi_\beta\Phi^\dag_\alpha\Phi_\gamma -
\Phi_\gamma\Phi^\dag_\alpha\Phi_\beta=0,\quad
\Tr(\Phi_\beta\Phi^\dag_\alpha) = \delta_{\alpha\beta}.
\end{equation}
However, for a nontrivial deformation, i.e.\,\,$\chi(N)\not\equiv
N$, the double and higher commutators (or anticommutators) are
significant and must be taken into account (we drop them).

\section{Composite Quasifermions\\ with Non-Deformed Constituent Boson} \label{sec:nd_gen}

Consider the realization of CFs formed of a usual boson and a
fermion.\,\,The respective wavefunctions of CFs realized by usual
fermions satisfy Eqs.~(\ref{nondef}) which will be solved
below.\,\,Let us choose a matrix $\Phi_\alpha = \Phi_{\alpha_1}$
with maximal rank and perform the singular value (SVD- or Schmidt-)
decomposition:
\[
\Phi_{\alpha_1}\! = U_1 D_{\alpha_1} V^\dag_1, \quad
\]
\[
 D_{\alpha_1}\! = \diag\{\lambda^{\alpha_1}_i\!, i =\kappa_1,...,\kappa_r,...\}=
\]\vspace*{-9mm}
\begin{equation}
=
   \diag\{\lambda^{\alpha_1}_{\kappa_1} E_{m_1},
        \lambda^{\alpha_1}_{\kappa_2} E_{m_2}, ...,
        \lambda^{\alpha_1}_{\kappa_r} E_{m_r}\}  \label{Phi_alpha1}
\end{equation}
with real non-negative $\lambda^{\alpha_1}_{\kappa_l}$ put in
descending order\,\footnote[3]{Physical meaning of
indices~$\kappa_l$ may be the relative momentum of constituents in
the c.m.\,\,system plus other quantum numbers.},
$\lambda^{\alpha_1}_{\kappa_1} \!>\! ...\,\,\!>\!
\lambda^{\alpha_1}_{\kappa_r} \!\ge\! 0$, obeying $\sum_l m_l
(\lambda^{\alpha_1}_{\kappa_l})^2 \!=\! 1$, and unitary matrices
$U_1$, $V_1$.\,\,For remaining matrices $\Phi_\gamma$, $\gamma\ne
\alpha_1$, we make replacement $\Phi_\gamma \to \tilde\Phi_\gamma$:
\begin{equation}   \label{repl}
\Phi_\gamma = U_1 \tilde\Phi_\gamma V^\dag_1, \ \ \ \gamma\ne \alpha_1.
\end{equation}
The first equation in~(\ref{nondef}) at $\alpha\!=\!\beta\!=\!\alpha_1$ now reads
\[
D_{\alpha_1}^2 \tilde\Phi_\gamma \!-\! \tilde\Phi_\gamma D_{\alpha_1}^2 \!=\! 0
\ \ {\rm or}\ \
\bigl((\lambda^{\alpha_1}_i)^2 \!-\! (\lambda^{\alpha_1}_j)^2\bigr)
(\tilde\Phi_\gamma)_{ij} \!=\! 0. 
\]
If $\lambda^{\alpha_1}_i\!\ne\! \lambda^{\alpha_1}_j,$ we have
$(\tilde\Phi_\gamma)_{ij} \!=\! 0$.\,\,Ac\-cor\-ding to the
block-diagonal form of $D_{\alpha_1}$, see~(\ref{Phi_alpha1}), the
other matrices~$\tilde\Phi_\gamma$ also take block-diagonal form:
\[
\tilde\Phi_\gamma = \diag\{\tilde\Phi_{\gamma,1}, \tilde\Phi_{\gamma,2},...,
 \tilde\Phi_{\gamma,r}\}, \ \ \gamma\ne \alpha_1.
\]
The dimensions of unit matrices $E_{m_k}$ and square matrices
$\tilde\Phi_{\alpha,k}$ are equal to the multiplicities of singular values
 $\lambda^{\alpha_1}_{\kappa_k}$.\,\,Now, the first equation in~(\ref{nondef})
reduces to the set of $r$ independent systems, $k=1,2,..,r$:
\begin{equation}
\lambda^{\alpha_1}_{\kappa_k} (\tilde{\Phi}^\dag_{\beta,k}
\tilde{\Phi}_{\gamma,k} - \tilde{\Phi}_{\gamma,k}
\tilde{\Phi}^\dag_{\beta,k}) =0,\label{phi-norm}
\end{equation}\vspace*{-9mm}
\begin{equation}
\lambda^{\alpha_1}_{\kappa_k} (\tilde{\Phi}_{\beta,k}
\tilde{\Phi}_{\gamma,k} - \tilde{\Phi}_{\gamma,k}
\tilde{\Phi}_{\beta,k}) =0,\label{phi-commut}
\end{equation}\vspace*{-9mm}
\begin{equation}
\tilde{\Phi}_{\beta,k} \tilde{\Phi}^\dag_{\alpha,k}
\tilde{\Phi}_{\gamma,k} \!-\! \tilde{\Phi}_{\gamma,k}
\tilde{\Phi}^\dag_{\alpha,k} \tilde{\Phi}_{\beta,k} =0,
    \ \ \alpha,\!\beta,\!\gamma\! \ne\! \alpha_1. \label{phi-tilde}
\end{equation}
From (\ref{phi-norm})--(\ref{phi-commut}), we infer that the
matrices~$\tilde\Phi_{\gamma,k}$, $\gamma\ne $ $\ne\alpha_1$,
constitute the set of commuting normal matrices (commuting with
their Hermitian conjugate ones) at fixed $k=1,2,..,r-1$ and possibly
at fixed $k=r$, if $\lambda^{\alpha_1}_{\kappa_r}\ne 0$.\,\,Un\-der
such premises, as known~\cite{Gantmacher2000ThMatrV1}, there is a
fixed unitary matrix~$\tilde{U}_k$ such that $\tilde\Phi_{\gamma,k}
= \tilde{U}_k \tilde{D}_{\gamma,k} \tilde{U}_k^\dag$ with diagonal
one~$\tilde{D}_{\gamma,k}$, $\gamma\ne \alpha_1$.\,\,If
$\lambda^{\alpha_1}_{\kappa_r} \ne 0,$ Eqs.~(\ref{phi-tilde}) are
automatically satisfied, and we have
\begin{equation} \label{gen_nd_sol}
\Phi_\alpha \!=\! U D_\alpha V^\dag,\ \ D_\alpha|
\mathop{=}\limits_{\alpha\ne \alpha_1}
\diag\{\tilde{D}_{\alpha,1},...,\tilde{D}_{\alpha,r}\}
\end{equation}
with $U \!=\! U_1 \diag\{\tilde{U}_k\}$, $V \!=\! V_1 \diag\{\tilde{U}_k\}$, $k\!=\!\overline{1,r}$.

If $\lambda^{\alpha_1}_{\kappa_r}= 0,$ Eqs.~(\ref{phi-tilde}) at
$k=\overline{1,r\!-\!1}$ are satisfied, while the remaining one, at
$k=r$, for $\tilde\Phi_{\alpha,r}$, $\alpha\ne \alpha_1$, can be
solved like above for $\Phi_\alpha$.\,\,Thus, by induction on the
number $\#\alpha \equiv m_{CF}$ of matrices (modes) $\Phi_\alpha,$
we can show that~(\ref{gen_nd_sol}) with arbitrary $U$ and $V$
presents the general solution
of~(\ref{nondef})\,\footnote[4]{Indeed, if $\#\alpha=1,$ it is just
SVD.\,\,Let (\ref{gen_nd_sol}) be valid for $(\#\alpha)-1$
modes.\,\,Then, at $\lambda^{\alpha_1}_{\kappa_r} = 0$, the
induction assumption is to be applied to~(\ref{phi-tilde}).}.

\begin{figure}
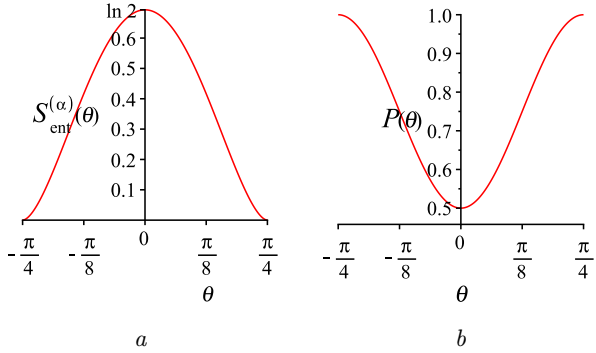

\vskip1mm
\includegraphics[width=3.9cm]{1a}\hspace{3mm}\includegraphics[width=3.9cm]{1b}\\
{\it a\hspace{4.1cm}b} \vskip-3mm \caption{Entanglement entropy
$S_{\rm ent}^{(\alpha)}$ ($a$) and purity $P$ versus the parameter
$\theta$ ($b$) for each mode~$\alpha=1,2$} \label{fig2}
\end{figure}

Conversely: all the matrices~(\ref{gen_nd_sol}) with $\Tr(D_\beta
\overline{D_\alpha}) =$ $= \delta_{\alpha\beta}$ satisfy
system~(\ref{nondef}) and thus give its general solution. The
diagonal elements of $D_\alpha$ are components of the orthonormal
vectors~$\lambda^{\alpha}$ in the complex space:
\begin{equation}
D_\alpha \!\!=\!
\diag\{\lambda^{\alpha}_1\!,\lambda^{\alpha}_2\!,...\}, \ \
(\lambda^\alpha \!\!,\! \lambda^\beta) \!\equiv\! \sum\limits_i
\!\!\lambda^{\alpha}_i \overline{\lambda^{\beta}_i} \!=\!
\delta_{\alpha\beta}. \!\!  \label{orth_nd}
\end{equation}
So, the ``boson-fermion'' entanglement entropy for realized CFs in
the $\alpha$th mode is given along with the orthogonality constraint
as
\begin{equation}\label{Sent_nd}
S_{\rm entang}^{(\alpha)} \!=\! -\sum\limits_i
|\lambda^{\alpha}_i|^2 \ln|\lambda^{\alpha}_i|^2, \ \
(\lambda^\alpha \!,\! \lambda^\beta) = \delta_{\alpha\beta}.
\end{equation}
Note that, for a parametrization of solution~(\ref{gen_nd_sol}), the one
of suitable $SU(n)$ can be used: this concerns unitary
matrices~$U$, $V$ and diagonal matrices~$D_\alpha$, since
their elements~$(\lambda^\alpha_i)$ constitute rows (columns)
of a unitary matrix, see~(\ref{orth_nd}).

\textit{Remark 1.}\,\,While for the realized composite bosons, the
block~$U_\alpha(m)$ in~(\ref{gen_solution}) is associated with the
$\alpha$th mode, yielding $m^2\!-\!1$ free \textit{real} parameters
in wa\-ve\-func\-tions (maximally, $m^2/2$-dimensional {\it complex}
space of states), the realized CFs admit more general
wa\-ve\-func\-tions.\,\,In\-deed, the {\it particular}
or\-tho\-nor\-mal CF wavefunctions
\[
\Phi_\alpha \!=\! U \diag\Bigl\{0..0, \underbrace{U_\alpha(m)
D^{(\alpha)}_m V_\alpha^\dag(m)}_{GL(m)}, 0..0\Bigr\} V^\dag
\]
already have avg.~$m^2\!-\!1$ free \textit{complex} parameters per
mode.

\section{Cofermions in Low-Mode Cases}\label{sec:2mod}

Consider, in more details, low-mode  subcases with CFs in two
modes~$\alpha=1,\,2$, and constituents -- upto three
modes.\,\,Re\-call that the general solution in the case of
nondeformed constituent boson was given in Sec.~\ref{sec:nd_gen}.

\subsection*{5(a). Two or three modes\\ of a constituent fermion
            $+$ \textit{usual} boson}

First, let both constituents be in two modes, i.e. $\mu,\nu =
\overline{1,2}$.\,\,The solution for realized wavefunctions reads
\begin{equation}\label{sol2*2nd}
\Phi_\alpha \!= e^{-i\tilde\alpha \psi} U \!\left(\!\!
\begin{array}{cc}
    \cos(\tilde\alpha \!+\! \theta) e^{i\phi} \!\! & \!0 \\
    0\! & \!\!\sin(\tilde\alpha \!+\! \theta) e^{-i\phi}
\end{array}
\!\right)\! V^\dag ,
\end{equation}
where $\tilde\alpha = \frac{\pi}{2} (\alpha-\alpha_c) =\mp {\pi\over
4}$, $\alpha_c=3/2$.\,\,Due to a low dimensionality of matrices, it
is convenient  to use the angle parametrization of $SU(2)$.\,\,There
can be other parametrization as well.

The {\it entanglement entropy within the CF}, realized
by a usual fermion, for each of the two modes is as follows:
\begin{equation}
S_{\rm ent}^{(\alpha)} \!\equiv\!  -\sum\nolimits_k
(\lambda^{\alpha}_k)^2 \ln (\lambda^{\alpha}_k)^2
\Bigl|_{\alpha=1,2} \!\!=\! S_2\Bigl({\pi\over 4} \!+\!
\theta\Bigr)\!,\label{S_2(a)}
\end{equation}
where $S_2(x) \equiv -\sin^2 x \ln\sin^2 x
    - \cos^2 x \ln \cos^2 x$.
This result is visualized in Fig.~\ref{fig2},~$a$.\,\,The maximum
$S_{\rm ent}\!=\! \ln 2$ may correspond to the most tightly bound
state of a realized CF.\,\,Clear\-ly, $S_{\rm ent}=0$ means the
opposite, i.e.\,\,the most loosely bound one.\,\,Ano\-ther
entanglement measure of a CF state, {\it purity} $P,$ is
\[
P|_{\alpha=1,2} \!\equiv\! \sum\nolimits_k\! (\lambda^{\alpha}_k)^4
    \!=\! \frac14 (3-\cos 4\theta).
\]
The purity ranges from $P=1/2$ (at $\theta=0$) to $P=1$ (at
$\theta=\pm \frac{\pi}{4}$), see~Fig.~\ref{fig2}~$b$.

{\it Remark 2.}\,\,Nonseparable CF states with \textit{fixed}
intermediate ($0<S_{\rm ent}<\ln 2$) entanglement entropy and two
respective wavefunctions~$\Phi_\alpha^{\mu\nu}$ are parametrized, in
total, by 6 independent real parameters.

\subsection*{\textit{Non-deformed} constituents in three modes}

In this case, we take $\alpha=\overline{1,2}$,
$\mu,\nu=\overline{1,3}$ so that $\Phi_\alpha^{\mu\nu}$ are some
$3\times3$-matrices.\,\,The general solution~(\ref{gen_nd_sol})
reads
\[
\Phi_\alpha = e^{-i \psi(\alpha)}\,
U \diag\{\lambda^{\alpha}_1,\lambda^{\alpha}_2,\lambda^{\alpha}_3\} V^\dag,
\quad U,V \in SU(3),
\]      
with  {\it complex} $\lambda^{\alpha}_k$, $k=\overline{1,3}$,
satisfying the orthonormality conditions~(\ref{orth_nd}), and $U, V$
not depending on~$\alpha$.

A parametrization of two orthonormal vectors~$(\lambda^1_k)$ and
$(\lambda^2_k)$ follows from the one of $SU(3)$, since the
rows/colums of matrices from $SU(3)$ separately form orthonormal
vectors.\,\,In\-deed, using the parametrization
from~\cite{Bronzan1988ParamSU3}, we obtain
($\alpha\!=\!1\leftrightarrow\alpha=2$) symmetric unified
expressions:
\begin{align}
&(\lambda^{\alpha}_k) =\nonumber   \\
&=\!\!\left(\!
\begin{aligned}
& e^{i\phi_1}\! \bigl(\sin\theta_1\! \cos\theta_2\!
    \cos(\tilde\alpha +\theta_3\!)\,
    -\\[-1mm]
&-\, \sin\theta_2\! \sin(\tilde\alpha +\theta_3\!)
    e^{i\phi_2} \!\bigr),\\
  &  \cos\theta_1 \cos(\tilde\alpha \!+\!\theta_3),\\
& e^{-i\phi_3}\! \bigl(\sin\theta_1\! \sin\theta_2\!
    \cos(\tilde\alpha+\theta_3\!)\,
    +\\[-1mm]
  &  +\, \cos\theta_2\! \sin(\tilde\alpha \!+\!\theta_3\!)
     e^{i\phi_2} \!\bigr)
\end{aligned}\!\right)\!\!,\label{3x3nd-lam}
\end{align}
where $0\le \theta_1,\theta_2,\theta_3\!+\!\pi/4 \le \pi/2$,
$0\le \phi_{1,2,3} \le 2\pi$.
The {\it entanglement entropy $S_{\rm ent}^{(\alpha)}(\theta_1,\theta_2,\theta_3,\phi_2)$
of a CF} in the $\alpha$th mode stems from~(\ref{Sent_nd})
with ``symmetrized'' squared absolute values of Schmidt coefficients:
\[
\bigl|\lambda^\alpha_{1,3}\bigr|^2 = \frac14 (1\!+\! \sin^2\theta_1)
\pm \frac14\, \frac{\cos^4\theta_1}{1\!+\! \sin^2\theta_1}
\frac{\sin 2(\theta_3^+ \!-\! \theta_3^-)}{\sin 2(\theta_3^+ \!+\!
\theta_3^-)}\, -
\]\vspace*{-6mm}
\[
-\, \frac12\, \frac{\sin 2\theta_3^\pm}{\sin 2(\theta_3^+ \!+\!
\theta_3^-)} \, \cos^2\theta_1 \cos 2\bigl(\theta_3 \!\mp\!
\theta_3^\mp \!+\! \tilde\alpha\bigr),
\]\vspace*{-5mm}
\begin{equation}
|\lambda^\alpha_2|^2 = \frac12 (1- \sin^2\theta_1)
 + \frac12\, \cos^2\theta_1 \cos 2(\theta_3 + \tilde\alpha),
\end{equation}
where two ``shift'' angles~$\theta_3^\pm$ replace~$\theta_2,\phi_2$,
while the upper or lower sign corresponds to the first or,
respectively, third Schmidt coefficient.\,\,The transition
$(\theta_2,\phi_2) \to (\theta_3^-,\theta_3^+)$ is given by the
unified formula
\[
\tg 2\theta_3^\pm = \pm \frac{2 \sin\theta_1
\tg(\theta_2 + \frac{\pi}{2} \delta_{\mp1,1}) \cos\phi_2}
{1 - \sin^2\theta_1 \tg^2(\theta_2 + \frac{\pi}{2} \delta_{\mp1,1})}.
\]
Note that the ``shift'' is counted from the parameter~$\theta_3$
which has a special role being directly \textit{linked} with
the mode number~$\alpha$, see~(\ref{3x3nd-lam}).

{\it Remark 3}.\,\,Parametrization \textit{asymmetric} under
$(\alpha\!=$ $=1) \leftrightarrow (\alpha=2)$,
see~\cite{Bronzan1988ParamSU3}, of the modes can also play a role.
From the physics viewpoint, it is possible, when the realization is
applied to a system with {\it ad hoc} asymmetry, for example, due to
an applied external asymmetric field or like for $s$-
vs.\,\,$p$-levels/modes of~CF.

\begin{figure}[h!]
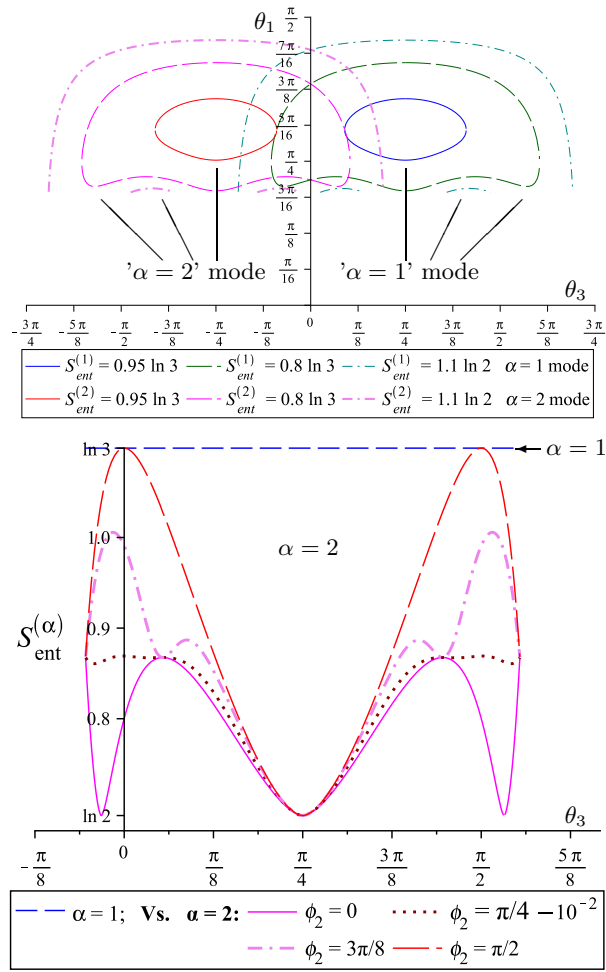

\centering \vskip-1mm
\begin{overpic}[width=\columnwidth]{fig3.eps}  
\put(92,22){$\theta_3$} \put(41,66){$\theta_1$}
\put(55,25){'$\alpha=1$' mode} 
\put(75,28){\line(1,1){10}}  
\put(70,28){\line(1,2){5}}
\put(66,28){\line(0,1){15}}  
\put(20,25){'$\alpha=2$' mode} \put(26,28){\line(-1,1){10}}
\put(31,28){\line(-1,2){5}} \put(35,28){\line(0,1){15}}
\end{overpic}\\
\begin{overpic}[width=\columnwidth]{fig3b.eps}
\put(92,27){$\theta_3$} \put(89,87.5){$\alpha=1$}
\put(88,88.5){\vector(-1,0){4}} \put(45,71){$\alpha=2$}
\end{overpic}
\vskip-4mm \caption{Upper: Equi-entropic curves ($S_{\rm
ent}^{(\alpha)} = \rm const$) versus angles $\theta_1$ and
$\theta_3$, at fixed~$\theta_3^\mp = {\pi\over3}$. Lower:
Entanglement entropy $S_{\rm ent}^{(2)}(\theta_3,\phi_2)$ for
cofermion in $\alpha=2$ mode at fixed entanglement entropy $S_{\rm
ent}^{(1)}\!=\!\max\!=\!\ln 3$ for $\alpha=1$ cofermion.}
\label{fig3}
\end{figure}

Thus, the {\it CF entanglement entropies}~$S_{\rm ent}^{(1)}$ and
$S_{\rm ent}^{(2)}$ are parametrized by three angles and one phase.
Unlike the constituents in the two-mode $\mu,\nu=\overline{1,2}$
case where $S_{\rm ent}^{(1)}-S_{\rm ent}^{(2)}=0$ and
$0\le S_{\rm ent}^{(\alpha)}\le \ln 2$, now, in the $\mu,\nu=\overline{1,3}$
case, we have $|S_{\rm ent}^{(1)}-S_{\rm ent}^{(2)}|\le \ln 2$,
while $0\le S_{\rm ent}^{(\alpha)}\le \ln 3$. The limits for above
differences impose a restriction on the realizable states.
 To illustrate the dependences $S_{\rm ent}^{(\alpha)} =
S_{\rm ent}^{(\alpha)}(\theta_1,\theta_3,\theta_3^\mp)$
at $C_3$-symmetric choice $\theta_3^\mp = {\pi\over3}$,
for both modes~$\alpha=1,2$, Fig.~\ref{fig3} (upper)
shows the equientropic curves versus $\theta_1$-, $\theta_3$-angles.
The somewhat similar behavior, though for the entropy of
mixing~\cite{Bolukbasi2006qutrits} within a three-level system, was
given in the context of the $SU(3)$ parametrization of qutrits
(see~\cite{Bolukbasi2006qutrits}).


Observe that, at $\theta_3^\pm = {\pi\over3}$ and $\sin^2\theta_1 = \frac13,$
the entropy for the $\alpha$th mode acquires
a $C_3$-cyclically symmetric form:
\begin{equation}
S_{\rm ent}^{(\alpha)} = \sum_{l=-1}^1  S_1\bigg(\!\frac23
\cos^2\bigl(\theta_3 + \frac{\pi}{3}l + \tilde\alpha\bigr)
\!\bigg)\!, \label{C3-cyclic_S}
\end{equation}
where $S_1(x) \!\equiv\! -x \ln x$.

\subsection*{5(b). CFs composed\\ of a fermion plus a \textit{deformed} boson} \label{sec:2-mod(def)}

Consider the two-mode case ($\alpha=1,\,2$) of CF composed of a
usual fermion and a $\chi$-{\it deformed} boson. The specificity of
two modes implies that it is sufficient to consider the realization
conditions~(\ref{req2'})--(\ref{req3}) on the ground and one-CF
states.\,\,Then, for the single non-zero two-CF state $A_1^\dag
A_2^\dag|0\rangle$, the realization conditions will be automatically
satisfied.\,\,Denoting $\delta\chi_2 \equiv \chi(2)\!-\!2$,
requirement~(\ref{req2_chi(2)}) on respective ``one-CF states''
reduces to two independent equations:
\[
L(\Phi_1,\Phi_2) \equiv \Phi_1\Phi_1^\dag\Phi_2 -
\Phi_2\Phi_1^\dag\Phi_1\, +
\]\vspace*{-9mm}
\begin{equation}\label{eq1def}
+\, \delta\chi_2 \bigl[\diag\{(\Phi_1\Phi_1^\dag)^{\mu\mu}\} \Phi_2
- \diag\{(\Phi_2\Phi_1^\dag)^{\mu\mu}\} \Phi_1\bigr] =0,
\end{equation}\vspace*{-9mm}
\begin{equation}
-L(\Phi_2,\Phi_1) =0.\label{eq2def}
\end{equation}
Using substitution~(\ref{repl}), restricting  to
$\mu,\nu\!=\!1,\,2$, \mbox{taking}\vspace*{-3mm}
\[
U_1 \!=\! U(u,v) \!\equiv\! \Bigl(\!
\begin{array}{cc}
    u & v \\
    -\overline{v} & \overline{u}
\end{array}
\!\Bigr) \!\in\! SU(2),~ |u|^2\!+\!|v|^2=1,
\]
 and applying the
identity\vspace*{-3mm}
\[
U^\dag \diag\{(U X U^\dag)^{\mu\mu}\} U= \frac12 X + \frac12 RXR,
\]
with the Hermitian
\[
R = \bigg(\!
\begin{array}{cc}
    |u|^2-|v|^2 & 2\overline{u}v \\
    2u\overline{v} & |v|^2-|u|^2
\end{array}\!\bigg)\!,
\]
we arrive at the matrix equations
\begin{equation}
\chi(2)\! (D_1^2 \tilde\Phi_2 \!-\! \tilde\Phi_2 D_1^2) \!+\!
    \delta\chi_2 R\bigl( D_1^2 R \tilde\Phi_2 \!-\! \tilde\Phi_2 D_1 R D_1\bigr) \!=\!0,
\end{equation}\vspace*{-10mm}
\[
\chi(2) (D_1 \tilde\Phi_2^\dag \tilde\Phi_2 - \tilde\Phi_2
\tilde\Phi_2^\dag D_1)\, +
\]\vspace*{-7mm}
\begin{equation}\label{eq1''def}
+\, \delta\chi_2 \bigl(R D_1 \tilde\Phi_2^\dag R \tilde\Phi_2
            - R \tilde\Phi_2 \tilde\Phi_2^\dag R D_1 \bigr) =0 .
\end{equation}
In view of three-dimensionality of the space
of~$\tilde{\Phi}^\dag_2$ satisfying the orthogonality condition
$\Tr(D_1\tilde{\Phi}^\dag_2) =0$, we present them as a linear
combination:
\[
\tilde{\Phi}_2 = x_1 \bigg(\!\!
\begin{array}{cc}
    \lambda^{(1)}_2 & 0 \\
    0 & -\lambda^{(1)}_1
\end{array}\!\!\bigg) +x_2 \bigg(\!\!
\begin{array}{cc}
    0 & \varkappa \lambda^{(1)}_1\\
    \overline{\varkappa}\lambda^{(1)}_2 & 0
\end{array}\!\!\bigg)\, +
\]\vspace*{-7mm}
\[
+\, x_3 \bigg(\!\!
\begin{array}{cc}
    0 & -\varkappa \lambda^{(1)}_2 \\
    \overline{\varkappa} \lambda^{(1)}_1 & 0
\end{array}\!\!\bigg)\!,
\]
where $\varkappa \!=\! e^{i(\arg v -\arg
u)}$.\,\,Equation~(\ref{eq1''def}) reduces to three linear (in
$x_1,x_2,x_3$) equations, the associated determinant should be zero:
\[
\det\bigl(...\bigr) = -\chi(2)(\chi(2)-2) |u|^2|v|^2
\bigl(|\lambda^{1}_1|^2\!-|\lambda^{1}_2|^2\bigr)^2 = 0.
\]
This is possible in the following cases:

{\bf a)} $\chi(2)=0$. The solution of (\ref{eq1def})--(\ref{eq2def})
reads
\begin{equation}\label{sol2*2d-a}
\Phi_\alpha \!= U(u,(-1)^{\alpha-\!1} v) \! \left(\!\!
\begin{array}{cc}
    \cos(\tilde\alpha + \theta) \!\! & \!0 \\
    0\! & \!\!\sin(\tilde\alpha + \theta)
\end{array}
\!\!\right)\! V^\dag.
\end{equation}

{\bf b)} $\chi(2)=1$. We find two classes of solutions:
\begin{equation}
\Phi_\alpha \!=  \! \diag\bigl\{\cos(\tilde\alpha + \theta),
            \sin(\tilde\alpha + \theta)\bigr\} V^\dag,      \label{sol2*2d-b1}
\end{equation}\vspace*{-7mm}
\begin{equation}
\Phi_\alpha \!\equiv \! (\Phi_\alpha^{\mu\nu}) =
\Bigl(\delta_{\mu\mu_0} \overline{V_{\nu\alpha}\!}\Bigr)\!,~~~
{\mathrm{fixed}} ~~\mu_0= 1~ {\mathrm{or}}~ 2. \label{sol2*2d-b2}
\end{equation}

{\bf c)} $\chi(2)=2,$ i.e. $\delta\chi_2=0$ -- non-deformed one,
see~(\ref{sol2*2nd}).

{\bf d)} At $\chi(2)\ne 0,1,2$, the solution is identical
to~(\ref{sol2*2d-b1}).

So, the {\it entanglement entropy of the cofermion} containing a
deformed boson is either given by the general parameter-dependent
expression, see~(\ref{S_2(a)}), or the constant $S_{\rm ent}=0$ in
the special case $\chi(2)= 1$, for~(\ref{sol2*2d-b2}).\,\,The
deformations $\chi(2)= 0$, $1,$ or $2$ are disjoint from the
continuous set.\,\,How the deformation parameter $\chi(2)$ is
reflected in physical quantities will be analyzed in the subsequent
paper.


\section{Conclusions} \label{sec:discussion}

After we settled the problem of realization of composite fermions
(CFs) by usual fermions, we explored the bipartite entanglement
(inside the CF) measured by the entanglement entropy of CF.\,\,The
analysis has been performed in relatively simpler cases: {\it i})
CFs with a non-deformed \textit{constituent boson}, for which we
have considered the examples of two and three modes for the both
constituents, {\it ii}) CF containing a {\it deformed} boson, both
in two modes.\,\,The resulting expressions are given
in~(\ref{Sent_nd}) in the case of non-deformed constituents,
(\ref{S_2(a)}) for CFs and constituents being in two modes, and
finally by~(\ref{3x3nd-lam}), (\ref{C3-cyclic_S}) for  two-mode CFs
with three-mode constituents.

   As is found, for the entanglement entropy of realized CFs of the type
``fermion + deformed boson'', the constituent boson deformation does
not manifest itself in the explicit formulas for the entanglement
entropy and purity of CF within each fixed subcase in these one- and
two-mode cases.\,\,At first sight, this is {\it in some contrast} to
the earlier studied entanglement entropy of
quasibosons~\cite{GM_Entang}, where the major issue was just the
dependence on the deformation parameter~$f$.\,\,The\-rein, all other
parameters of the states [like those from $U_1(d_a)$, $U_2(d_b)$,
$U_{\alpha}(m)$ in~(\ref{gen_solution})] did not enter the
entanglement measures.\,\,In the present case of CFs, the situation
is different: the CFs are realized by {\it nondeformed} fermions,
so, the analogous deformation parameter corresponding to CF as a
whole is absent.\,\,The quantity~$\chi(2),$ being related to the
deformation parameter(s), in the case of CF has different origin,
since it concerns the constituent of CF.\,\,Ne\-ver\-the\-less,
there appear additional parameters present in the matrix $\Phi$ of
ansatz (\ref{ansatz}) which, along with~$\chi(2)$, define the
entanglement entropy of CFs. Thus, these parameters determine the
form of CF states (their wavefunctions).\,\,This dependence on the
involved parameter is shown in Fig.~\ref{fig2}.\,\,Al\-so noteworthy
are the properties of CF entanglement entropy shown in
Fig.~\ref{fig3}.\,\,In that case, the behavior is apparently richer.

Let us note again that this paper presents
explicit formulas for the
entanglement entropy {\it inside an individual composite fermion}
 (i.e.\,\,for the interconstituent entanglement), see also Introduction.\,\,In contrast,
 the authors in~\cite{Gioev2006Entanglement,Shao2014Entanglement}
 explored the (statistical) entanglement entropy of {\it many-fermion systems}
 that occupy a certain space region.\,\,In particular, the efficient
 numerical methods were applied in~\cite{Shao2014Entanglement}
 to the system of 37 composite fermions, and the linear size
 of a subsystem entered the final result for the entanglement entropy.\,\,While the results of~\cite{Gioev2006Entanglement} depend
explicitly on the space region dimensionality, we
operate here with one or more modes irrespectively of a particular space dimensionality.

What about the role of a deformation parameter~$f$ in the situation
with quasibosons? In that case, we had~\cite{GM_Entang,GM_Ent(En)}
quite natural feature: the entanglement entropy was rising with
decreasing values of~$f$, i.e.\,\,with approaching the truly bosonic
behavior, either for the Fock states at a fixed mode or for the
coherent states.\,\,In the present case of CFs, the physical meaning
of the parameter(s) which the entanglement entropy and purity depend
upon is not clear enough, and that issue deserves the further
study.\,\,Ne\-ver\-the\-less, concerning the considered cases of 2
or 3 modes for the constituents, we may remark the
following.\,\,Since the only parameters affecting the intercomponent
entanglement of CF are $\theta$ (the 2-mode case,
see~(\ref{S_2(a)})) or $\theta_i$, $i=\overline{1,3}$, and $\phi_2$
(the 3-mode case, see~(\ref{3x3nd-lam})), they should correspond to
internal quantum numbers of CF like spin, parameter(s) of the
binding energy of CF, \textit{etc}.

Remark also that the above parameters (like $\theta$) of the
realized states can be related to such (rather unexpected)
parameters as CF constituents' mass ratio or reduced mass.\,\,That
concerns, e.g., the trion CF composed of an exciton, modeled by a
deformed boson, and an electron or a hole. It is motivated, say, by
Fig.~2 in~\cite{Spink_Trion}, where the trion binding energy depends
on the reduced mass of the electron-hole pair, while the extent of
bipartite entanglement usually is related to the binding energy of a
composite particle.\,\,We intend to explore such entanglement-energy
relation and other implications elsewhere.


\vskip3mm \textit{This work was partly supported by the National
Academy of Sciences of Ukraine (project No.\,\,0117U000237).}





\vspace*{3mm} \rezume{О.М.\,Гаврилик, Ю.А.\,Міщенко} {СКЛАДЕНІ
ФЕРМІОНИ\\ ЯК ДЕФОРМОВАНІ ОСЦИЛЯТОРИ:\\
ХВИЛЬОВІ ФУНКЦІЇ ТА ЗАПЛУТАНІСТЬ} {Складена структура частинок дещо
змінює їх статистику порівняно із класичними бозе- та
фермі-статистиками. Теорема про зв'язок спіну зі статистикою, отже,
не виконується. Скажімо, $\pi$-мезони, екситони, куперівські пари не
є ідеальними бозонами і, подібним чином, баріони не є простими
ферміонами. У попередніх статтях ми вивчали двочастинкові
\textit{складені бозони} (тобто квазібозони) за допомогою реалізації
їх через деформовані осцилятори. Були знайдені такі характеристики
міжкомпонентної заплутаності як ентропія заплутаності та чистота
(\textit{purity}) в термінах параметра деформації.
    У цій роботі ми виконуємо аналогічний розгляд складених частинок
\textit{фермі-типу} та досліджуємо їх у двох основних випадках: (i)
складені ферміони (чи коферміони, чи СФ-ни) типу ``бозон +
ферміон''; (ii) СФ-ни типу ``деформований бозон + ферміон''. Як ми
показуємо, коферміони, в обох випадках, допускають реалізацію лише
звичайними ферміонами. Випадок~(i) розглянуто повністю та знайдено
хвильові функції разом із мірами заплутаності. Випадок~(ii)
розглянуто в межах декількох мод, як для СФ-нів так і для складових.
Ентропію заплутаності та ``п'юріті'' визначено через задіяні
параметри і проілюстровано \mbox{графічно.}}

\end{document}